\documentclass[aps,prl,twocolumn,superscriptaddress,showpacs]{revtex4}
\usepackage[dvips]{graphicx}
\usepackage{bm,color}
\usepackage{dcolumn}

\def\lamb#1#2{$^{#1}_{\Lambda}${#2}} 
\def\lam#1#2{$^{#1}_{~\Lambda}${#2}} 
\def\etal{\textit{et al.}}

\begin{document}

\title {High Resolution Spectroscopy of $^{\bm {16}}_{\bm{~\Lambda}}$N by
Electroproduction}

\author{F.~Cusanno}
\affiliation{Istituto Nazionale di Fisica Nucleare, Sezione di Roma,
Piazzale A. Moro 2, I-00185 Rome, Italy}

\author{G.M.~Urciuoli}
\affiliation{Istituto Nazionale di Fisica Nucleare, Sezione di Roma,
Piazzale A. Moro 2, I-00185 Rome, Italy}

\author{A.~Acha}
\affiliation{Florida International University, Miami, Florida 33199, USA}

\author{P.~Ambrozewicz}
\affiliation{Florida International University, Miami, Florida 33199, USA}

\author{K.A.~Aniol}
\affiliation{California State University, Los Angeles, Los Angeles
California 90032, USA}

\author{P.~Baturin}
\affiliation{Rutgers, The State University of New Jersey, Piscataway,
New Jersey 08855, USA}

\author{P.Y.~Bertin}
\affiliation{Universit\'{e} Blaise Pascal/IN2P3, F-63177 Aubi\`{e}re, France}

\author{H.~Benaoum}
\affiliation{Syracuse University, Syracuse, New York 13244, USA}

\author{K.I.~Blomqvist}
\affiliation{Universit\"at Mainz, Mainz, Germany}

\author{W.U.~Boeglin}
\affiliation{Florida International University, Miami, Florida 33199, USA}

\author{H.~Breuer}
\affiliation{University of Maryland, College Park, Maryland 20742, USA}

\author{P.~Brindza}
\affiliation{Thomas Jefferson National Accelerator Facility, Newport News,
Virginia 23606, USA}

\author{P.~Byd\v{z}ovsk\'y}
\affiliation{Nuclear Physics Institute, \v{R}e\v{z} near Prague, Czech
Republic}

\author{A.~Camsonne}
\affiliation{Universit\'{e} Blaise Pascal/IN2P3, F-63177 Aubi\`{e}re, France}

\author{C.C.~Chang}
\affiliation{University of Maryland, College Park, Maryland 20742, USA}

\author{J.-P.~Chen}
\affiliation{Thomas Jefferson National Accelerator Facility, Newport News,
Virginia 23606, USA}

\author{Seonho~Choi}
\affiliation{Temple University, Philadelphia, Pennsylvania 19122, USA}

\author{E.A.~Chudakov}
\affiliation{Thomas Jefferson National Accelerator Facility, Newport News,
Virginia 23606, USA}

\author{E.~Cisbani}

\author{S.~Colilli}
\affiliation{Istituto Nazionale di Fisica Nucleare, Sezione di Roma, gruppo
collegato  Sanit\`a, and Istituto Superiore di Sanit\`a, I-00161 Rome, Italy}

\author{L.~Coman}
\affiliation{Florida International University, Miami, Florida 33199, USA}

\author{B.J.~Craver}
\affiliation{University of Virginia, Charlottesville, Virginia 22904, USA}

\author{G.~De~Cataldo}
\affiliation{Istituto Nazionale di Fisica Nucleare, Sezione di Bari and
University of Bari, I-70126 Bari, Italy}

\author{C.W.~de~Jager}
\affiliation{Thomas Jefferson National Accelerator Facility, Newport News,
Virginia 23606, USA}

\author{R.~De~Leo}
\affiliation{Istituto Nazionale di Fisica Nucleare, Sezione di Bari and
University of Bari, I-70126 Bari, Italy}

\author{A.P.~Deur}
\affiliation{University of Virginia, Charlottesville, Virginia 22904, USA}

\author{C.~Ferdi}
\affiliation{Universit\'{e} Blaise Pascal/IN2P3, F-63177 Aubi\`{e}re, France}

\author{R.J.~Feuerbach}
\affiliation{Thomas Jefferson National Accelerator Facility, Newport News,
Virginia 23606, USA}

\author{E.~Folts}
\affiliation{Thomas Jefferson National Accelerator Facility, Newport News,
Virginia 23606, USA}

\author{R.~Fratoni}

\author{S.~Frullani}

\author{F.~Garibaldi}
\affiliation{Istituto Nazionale di Fisica Nucleare, Sezione di Roma, gruppo
collegato  Sanit\`a, and Istituto Superiore di Sanit\`a, I-00161 Rome, Italy}

\author{O.~Gayou}
\affiliation{Massachussets Institute of Technology, Cambridge, Massachusetts
02139, USA}

\author{F.~Giuliani}
\affiliation{Istituto Nazionale di Fisica Nucleare, Sezione di Roma, gruppo
collegato  Sanit\`a, and Istituto Superiore di Sanit\`a, I-00161 Rome, Italy}

 \author{J.~Gomez}
 \affiliation{Thomas Jefferson National Accelerator Facility, Newport News,
Virginia 23606, USA}

\author{M.~Gricia}
\affiliation{Istituto Nazionale di Fisica Nucleare, Sezione di Roma, gruppo
collegato  Sanit\`a, and Istituto Superiore di Sanit\`a, I-00161 Rome, Italy}

\author{J.O.~Hansen}
\affiliation{Thomas Jefferson National Accelerator Facility, Newport News,
Virginia 23606, USA}

\author{D.~Hayes}
\affiliation{Old Dominion University, Norfolk, Virginia 23508, USA}

\author{D.W.~Higinbotham}
\affiliation{Thomas Jefferson National Accelerator Facility, Newport News,
Virginia 23606, USA}

\author{T.K.~Holmstrom}
\affiliation{College of William and Mary, Williamsburg, Virginia 23187, USA}

\author{C.E.~Hyde}
\affiliation{Old Dominion University, Norfolk, Virginia 23508, USA}
\affiliation{Universit\'{e} Blaise Pascal/IN2P3, F-63177 Aubi\`{e}re, France}

\author{H.F.~Ibrahim}
\affiliation{Old Dominion University, Norfolk, Virginia 23508, USA}

\author{M.~Iodice}
\affiliation{Istituto Nazionale di Fisica Nucleare, Sezione di Roma Tre,
I-00146 Rome, Italy}

\author{X.~Jiang}
\affiliation{Rutgers, The State University of New Jersey, Piscataway,
New Jersey 08855, USA}

\author{L.J.~Kaufman}
\affiliation{University of Massachussets Amherst, Amherst,  Massachusetts
01003, USA}

\author{K.~Kino}
\affiliation{Research Center for Nuclear Physics, Osaka
University, Ibaraki, Osaka 567-0047, Japan}

\author{B.~Kross}
\affiliation{Thomas Jefferson National Accelerator Facility, Newport News,
Virginia 23606, USA}

\author{L.~Lagamba}
\affiliation{Istituto Nazionale di Fisica Nucleare, Sezione di Bari and
University of Bari, I-70126 Bari, Italy}

\author{J.J.~LeRose}
\affiliation{Thomas Jefferson National Accelerator Facility, Newport News,
Virginia 23606, USA}

\author{R.A.~Lindgren}
\affiliation{University of Virginia, Charlottesville, Virginia 22904, USA}

\author{M.~Lucentini}
\affiliation{Istituto Nazionale di Fisica Nucleare, Sezione di Roma, gruppo
collegato  Sanit\`a, and Istituto Superiore di Sanit\`a, I-00161 Rome, Italy}

\author{D.J.~Margaziotis}
\affiliation{California State University, Los Angeles, Los Angeles
California 90032, USA}

\author{P.~Markowitz}
\affiliation{Florida International University, Miami, Florida 33199, USA}

\author{S.~Marrone}
\affiliation{Istituto Nazionale di Fisica Nucleare, Sezione di Bari and
University of Bari, I-70126 Bari, Italy}

\author{Z.E.~Meziani}
\affiliation{Temple University, Philadelphia, Pennsylvania 19122, USA}

\author{K.~McCormick}
\affiliation{Rutgers, The State University of New Jersey, Piscataway,
New Jersey 08855, USA}

\author{R.W.~Michaels}
\affiliation{Thomas Jefferson National Accelerator Facility, Newport News,
Virginia 23606, USA}

\author{D.J.~Millener}
\affiliation{Brookhaven National Laboratory, Upton, New York 11973, USA}

\author{T.~Miyoshi}
\affiliation{Tohoku University, Sendai, 980-8578, Japan}

\author{B.~Moffit}
\affiliation{College of William and Mary, Williamsburg, Virginia 23187, USA}

\author{P.A.~Monaghan}
\affiliation{Massachussets Institute of Technology, Cambridge, Massachusetts
02139, USA}

\author{M.~Moteabbed}
\affiliation{Florida International University, Miami, Florida 33199, USA}

\author{C.~Mu\~noz~Camacho}
\affiliation{CEA Saclay, DAPNIA/SPhN, F-91191 Gif-sur-Yvette, France}

\author{S.~Nanda}
\affiliation{Thomas Jefferson National Accelerator Facility, Newport News,
Virginia 23606, USA}

\author{E.~Nappi}
\affiliation{Istituto Nazionale di Fisica Nucleare, Sezione di Bari and
University of Bari, I-70126 Bari, Italy}

\author{V.V.~Nelyubin}
\affiliation{University of Virginia, Charlottesville, Virginia 22904, USA}

\author{B.E.~Norum}
\affiliation{University of Virginia, Charlottesville, Virginia 22904, USA}

\author{Y.~Okasyasu}
\affiliation{Tohoku University, Sendai, 980-8578, Japan}

\author{K.D.~Paschke}
\affiliation{University of Massachussets Amherst, Amherst,  Massachusetts
01003, USA}

\author{C.F.~Perdrisat}
\affiliation{College of William and Mary, Williamsburg, Virginia 23187, USA}

\author{E.~Piasetzky}
\affiliation{School of Physics and Astronomy, Sackler Faculty of Exact
Science, Tel Aviv University, Tel Aviv 69978, Israel}

\author{V.A.~Punjabi}
\affiliation{Norfolk State University, Norfolk, Virginia 23504, USA}

\author{Y.~Qiang}
\affiliation{Massachussets Institute of Technology, Cambridge, Massachusetts
02139, USA}

\author{B.~Raue}
\affiliation{Florida International University, Miami, Florida 33199, USA}

\author{P.E.~Reimer}
\affiliation{Argonne National Laboratory, Argonne, Illinois 60439, USA}

\author{J.~Reinhold}
\affiliation{Florida International University, Miami, Florida 33199, USA}

\author{B.~Reitz}
\affiliation{Thomas Jefferson National Accelerator Facility, Newport News,
Virginia 23606, USA}

\author{R.E.~Roche}
\affiliation{Florida State University, Tallahassee, Florida 32306, USA}

\author{V.M.~Rodriguez}
\affiliation{University of Houston, Houston, Texas 77204, USA}

\author{A.~Saha}
\affiliation{Thomas Jefferson National Accelerator Facility, Newport News,
Virginia 23606, USA}

\author{F.~Santavenere}
\affiliation{Istituto Nazionale di Fisica Nucleare, Sezione di Roma, gruppo
collegato  Sanit\`a, and Istituto Superiore di Sanit\`a, I-00161 Rome, Italy}

\author{A.J.~Sarty}
\affiliation{St. Mary's University, Halifax, Nova Scotia, Canada}

\author{J.~Segal}
\affiliation{Thomas Jefferson National Accelerator Facility, Newport News,
Virginia 23606, USA}

\author{A.~Shahinyan}
\affiliation{Yerevan Physics Institute, Yerevan, Armenia}

\author{J.~Singh}
\affiliation{University of Virginia, Charlottesville, Virginia 22904, USA}

\author{S.~\v{S}irca}
\affiliation{Dept. of Physics, University of Ljubljana, Slovenia}

\author{R.~Snyder}
\affiliation{University of Virginia, Charlottesville, Virginia 22904, USA}

\author{P.H.~Solvignon}
\affiliation{Temple University, Philadelphia, Pennsylvania 19122, USA}

\author{M.~Sotona}
\affiliation{Nuclear Physics Institute, \v{R}e\v{z} near Prague, Czech
Republic}

\author{R.~Subedi}
\affiliation{Kent State University, Kent, Ohio 44242, USA}

\author{V.A.~Sulkosky}
\affiliation{College of William and Mary, Williamsburg, Virginia 23187, USA}

\author{T.~Suzuki}
\affiliation{Tohoku University, Sendai, 980-8578, Japan}

\author{H.~Ueno}
\affiliation{Yamagata University, Yamagata 990-8560, Japan}

\author{P.E.~Ulmer}
\affiliation{Old Dominion University, Norfolk, Virginia 23508, USA}

\author{P.~Veneroni}
\affiliation{Istituto Nazionale di Fisica Nucleare, Sezione di Roma, gruppo
collegato  Sanit\`a, and Istituto Superiore di Sanit\`a, I-00161 Rome, Italy}

\author{E.~Voutier}
\affiliation{LPSC, Universit\'e Joseph Fourier, CNRS/IN2P3, INPG, F-38026
Grenoble, France}

\author{B.B.~Wojtsekhowski}
\affiliation{Thomas Jefferson National Accelerator Facility, Newport News,
Virginia 23606, USA}

\author{X.~Zheng}
\affiliation{Argonne National Laboratory, Argonne, Illinois 60439, USA}

\author{C.~Zorn}
\affiliation{Thomas Jefferson National Accelerator Facility, Newport News,
Virginia 23606, USA}

\collaboration{Jefferson Lab Hall A Collaboration}
\noaffiliation

\date{\today}

 \begin{abstract}
An experimental study of the $^{16}$O$(e,e'K^+)$\lam{16}{N} reaction
has been performed at Jefferson Lab. A thin film of falling 
water was used as a target. This permitted a simultaneous measurement
of the p$(e,e'K^+)\Lambda$,$\Sigma^0$ exclusive reactions and a precise
calibration of the energy scale. A ground-state binding energy of 
$13.76\pm 0.16$ MeV was obtained for \lam{16}{N} with better
precision than previous measurements on the
mirror hypernucleus \lam{16}{O}. Precise energies have been determined
for peaks arising from a $\Lambda$ in $s$ and $p$ orbits coupled to
the $p_{1/2}$ and  $p_{3/2}$ hole states of the $^{15}$N core
nucleus.  
\end{abstract}


\maketitle

 Hypernuclear spectroscopy is a powerful tool for studying
the $\Lambda N$ interaction ($N\!=\!p, n$) given that very limited information
can be obtained from  $Y p$ ($Y\!=\!\Lambda,\Sigma^-,\Sigma^+$)
elastic scattering and charge-exchange data~\cite{rijken99}. 

 Information on the spin-dependent part of the $\Lambda N$ interaction
can be extracted from the energy splitting of hypernuclear spin doublets
formed when a $\Lambda$ in an $s$ state couples to a nuclear core
state with non-zero spin. Seven such doublet splittings have been
measured in \lamb{7}{Li}, \lamb{9}{Be}, \lam{11}{B}, 
\lam{15}{N}, and \lam{16}{O} using an array of germanium detectors called 
the Hyperball \cite{hashtam06,tamura08,ukai04,ukai08} and interpreted in 
terms of shell-model calculations
that include both $\Lambda$ and $\Sigma$ hypernuclear configurations
\cite{millener08}. For \lam{16}{O}, the ground-state is found to
have $J^\pi\!=\!0^-$ with excited states at 26 keV ($1^-$), 6562 keV
($1^-$), and 6782 keV ($2^-$) \cite{ukai04,ukai08} (the $2^-$ state
is weakly populated and its existence less certain).

 More details of the effective $\Lambda N$ interaction can in principle 
be obtained when the $\Lambda$ is in a $p$ state. With the exception of a
$p_{1/2\Lambda}$/$p_{3/2\Lambda}$ doublet in \lam{13}{C}~\cite{kohri02},
such states in p-shell nuclei are above particle thresholds and
cannot be studied via $\gamma$-ray spectroscopy. Thus, experimental studies
up until now have been carried out by hadron-induced reactions
with limited energy resolution.

In fact, $^{16}$O targets have been extensively used in hypernuclear
studies with the $(K^-,\pi^-)$, $(\pi^+, K^+)$, and
$(K^-_\mathrm{stop},\pi^-)$ reactions with dominant non-spin-flip
reaction mechanisms that excite natural-parity states \cite{hashtam06}. 
In all cases, four peaks are seen with the excited states at
$\approx 6.2$, $\approx 10.6$, and $\approx 17.1$ MeV corresponding
to $\Lambda$'s in $s$ and $p$ orbits coupled to the $p^{-1}_{1/2}$ 
ground state and the 6.176-MeV $p^{-1}_{3/2}$ states of $^{15}$O.
In the simple particle-hole limit, the degenerate multiplets contain 
2, 2, 4, and 6 states, respectively, and the cross sections would
be in the ratio 2:1 for peaks based on the $p_{3/2}$ vs. $p_{1/2}$
hole states. The first two peaks correspond to $1^-$ states
and the $B_\Lambda$ value for the lowest $1^-$ state is not 
particularly well determined. In the CERN $(K^-,\pi^-)$ experiment 
\cite{bruckner78}, the third and fourth peaks correspond to 
substitutional $0^+$ states. At the larger momentum transfer of the 
stopped $K^-$ work at KEK \cite{tamura94}, the same peaks contain 
contributions from both $0^+$ and $2^+$ hypernuclear states. Finally, 
in the $(\pi^+, K^+)$ reaction, first performed at BNL \cite{pile91} 
and later at KEK \cite{hashtam06} with better energy 
resolution, only the $2^+$ states are expected to contribute.

 The experimental knowledge can be enhanced using the $(e,e'K^+)$
electroproduction reaction characterized  by a large momentum transfer
to the hypernucleus ($q \gtrsim$ 250 MeV/c) and strong spin-flip terms,
even at zero $K^+$ production angles, resulting in the excitation
of both natural- and unnatural-parity states. In the present case,
$1^-$, $2^-$, $1^+$, $2^+$, and $3^+$ particle-hole states can be
excited with significant cross sections. In addition, the
$K^+\Lambda$ associated production occurs on a proton making
\lam{16}{N}, the mirror to \lam{16}{O}. After taking into account 
that the $p_{3/2}$-hole state is 148 keV higher in $^{15}$N than 
$^{15}$O, comparison of the energy spectra (and especially of 
$\Lambda$ binding energies) of these mirror hypernuclei can, 
in principle, shed light on charge-dependent effects in 
hyperon-nucleon interactions.

 The E94-107 experiment in Hall A at Jefferson Lab \cite{proposal}
started a systematic study of high resolution hypernuclear
spectroscopy in the $1p$ shell region of nuclei, such as $^{9}$Be,
$^{12}$C, and $^{16}$O. Results on $^{12}$C have been published 
\cite{iodice07} and the results on $^{16}$O are presented in this paper.

Hall A at JLab is well suited to perform $(e,e'K^+)$ experiments. 
Scattered electrons are detected in the High Resolution Spectrometer
(HRS) electron arm while coincident kaons are detected in the HRS hadron 
arm~\cite{nimhalla}. The disadvantage of smaller electromagnetic cross 
sections is partially compensated for by the high current and high duty 
cycle capabilities of the beam.
Throughout the experiment, the same equipment has been used in very
similar kinematical conditions on C, Be, and H$_2$O targets.
The use of a pair of septum magnets permitted particle detection at very
forward angles \cite{septum} and a Ring Imaging CHerenkov (RICH) detector
\cite{rich2004,pylosfg,pylosfc} has been used in the hadron arm to
provide an unambiguous identification of kaons when combined with 
the stardard particle identification apparatus of Hall A, based on aerogel 
Cherenkov detectors \cite{perrino,lagamba,marrone}.
Details and motivations 
for the specific choices can be found in Ref.~\cite{iodice07}.
In the case of the waterfall target, the kinematics were set to particle 
detection at  $6^\circ$ for both electrons and kaons, incident beam energy 
of 3.66 GeV, scattered electron momentum of 1.45 GeV/c, and kaon momentum 
of 1.96 GeV/c. The electron beam current was about 60 $\mu$A. The
spread of the beam energy was $< 3\times 10^{-5}$, continuously
monitored with a synchrotron-light interferometer \cite{chevtsov03}.

In the present experiment, a waterfall target \cite{waterfall} 
was used because pure oxygen is difficult to handle and the use of other
oxygen compounds requires additional measurements to subtract the non-oxygen
background. The presence of the hydrogen has many advantages. In particular, 
it permits a calibration of the missing-mass scale and thus an accurate
measurement of the $\Lambda$ binding energy in the hypernucleus.
Moreover, an interesting measurement of the $(e,e'K^+)$ cross section
in a previously unmeasured kinematical region was possible.

\begin{figure}
\centering
\includegraphics[width=8.0cm]{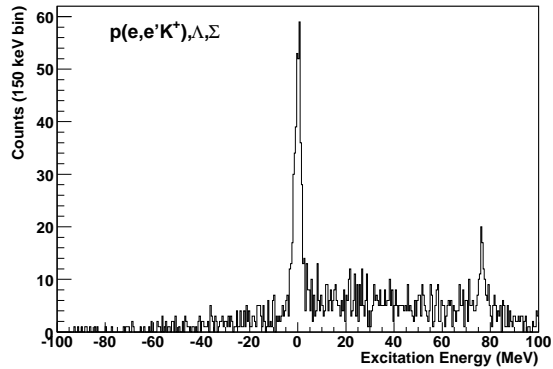}
\caption{Excitation energy spectrum of the $p(e,e'K^+)\Lambda,\Sigma^0$ 
on hydrogen, used for energy scale calibration. The fitted positions
(not shown on the plot) for the peaks are $-0.04\pm 0.08$ MeV and
$76.33\pm 0.24$ MeV.}
\label{fig:calib}
\end{figure}

\begin{figure}
\centering
\includegraphics[width=8.0cm, height=6.0cm]{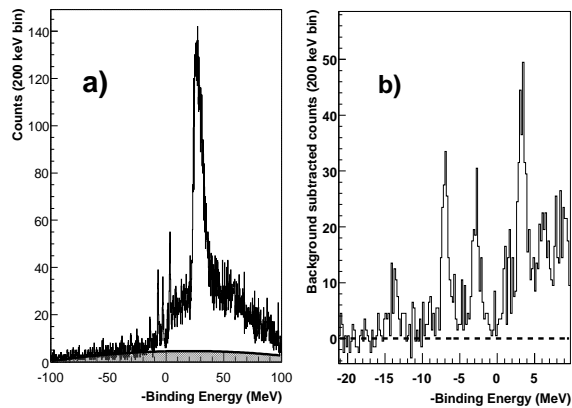}
\caption{The \lam{16}{N} binding-energy spectrum 
obtained after
kaon selection with aerogel detectors and RICH. The electron-kaon random
coincidence contribution evaluated in a large timing window is superimposed
on the spectrum in the left panel. The right panel shows the spectrum after
this background has been subtracted.}
\label{oxygen1}
\end{figure}

A complete calibration of the target thickness as a function of pump 
speed has been performed, the thickness was determined from 
the elastic cross section on hydrogen~\cite{waterfall}. The target 
thickness used was $75\!\pm\!3\,(\mathrm{stat.})\pm\!12\,(\mathrm{syst.})$
mg/cm$^2$. To calibrate the energy scale, the $\Lambda$ peak position
from the reaction on hydrogen was first obtained using the nominal 
central values for the  kinematic variables, and then constrained
to be zero by applying a small shift to the energy of the beam
(the quantity with the largest uncertainty). This shift is
common to reactions on hydrogen and oxygen and therefore its uncertainty
does not affect the determination of the binding energies of the
\lam{16}{N} levels. A resolution of 800 keV FWHM for the $\Lambda$ peak on
hydrogen is obtained. The linearity of the scale has been verified
from the $\Sigma^0 -\Lambda$ mass difference of 76.9 MeV. For
this purpose, a few hours of calibration data were taken with a slightly lower
kaon momentum (at fixed angles) to have the $\Lambda$ and $\Sigma^0$
peaks within the detector acceptance. Fig.~\ref{fig:calib} shows the two peaks 
associated with p$(e,e'K^+)\Lambda$ and p$(e,e'K^+)\Sigma^0$ production.
The linearity is verified to 
$(76.9 - 76.4\pm 0.3)/76.4 = 0.65\pm 0.40\%$,

\begin{table*}
\caption{Excitation energies, widths, and cross sections obtained by 
fitting the $^{16}$O$(e,e'K^+)$\lam{16}{N} spectrum (first three columns)
compared with theoretical predictions (last four columns). The excitation energies of the
remaining $p_N^{-1}p_\Lambda$ states, weakly populated by the
$(e,e'K^+)$ reaction, are 11.17 ($0^+$), 11.27 ($1^+$), 17.21 ($2^+$).
17.27 ($1^+$), 17.67 ($0^+$), and 18.35 ($1^+$) MeV; note that
spin-flip excitation for a $p_{3/2N}\to p_{3/2\Lambda}$ $2^+$ 
transition is forbidden (17.21-MeV state).
\label{tab:results}}
\begin{ruledtabular}
\begin{tabular}{ccccccc}
 & \multicolumn{2}{c}{Experimental data}  & &  \multicolumn{2}{c}{Theoretical prediction} \\ 
$E_x$ &  Width  &  Cross section & $E_x$  & Wave
function & $J^\pi$ & Cross section \\
 (MeV)  &  (FWHM, MeV) & $(nb/sr^2/GeV)$ & (MeV) &   &   &
 $(nb/sr^2/GeV)$\\
 \hline
 \phantom{1}0.00\phantom{ $\pm$ 0.06}   & 1.71 $\pm$ 0.70 & 1.45 $\pm$ 0.26 & 0.00 
& $p^{-1}_{1/2}\otimes s_{1/2\Lambda}$& $0^-$ & 0.002 \\
   &  &  & 0.03 & $p^{-1}_{1/2}\otimes s_{1/2\Lambda}$ & $1^-$ & 1.45 \\
 & & & & & & \\
\phantom{1}6.83 $\pm$ 0.06 & 0.88 $\pm$ 0.31 & 3.16 $\pm$ 0.35 & 6.71 &
$p^{-1}_{3/2}\otimes s_{1/2\Lambda}$ & $1^-$ & 0.80 \\
   &  &  & 6.93 &$p^{-1}_{3/2}\otimes s_{1/2\Lambda}$ & $2^-$   &2.11 \\
 & & & & & & \\
10.92 $\pm$ 0.07 & 0.99 $\pm$ 0.29 & 2.11 $\pm$ 0.37 & 11.00 &
$p^{-1}_{1/2}\otimes p_{3/2\Lambda}$ & $2^+$ & 1.82 \\
  & & & 11.07 & $p^{-1}_{1/2}\otimes p_{1/2\Lambda}$ & $1^+$ & 0.62 \\
 & & & & & & \\
17.10 $\pm$ 0.07 & 1.00 $\pm$ 0.23 & 3.44 $\pm$ 0.52 & 17.56 & 
$p^{-1}_{3/2}\otimes p_{1/2\Lambda}$ & $2^+$ & 2.10 \\
  & &  & 17.57 &
$p^{-1}_{3/2}\otimes p_{3/2\Lambda}$ & $3^+$ & 2.26 \\
\end{tabular}
\end{ruledtabular}
\end{table*}

\begin{figure}
\centering
 \includegraphics[width=8.0cm]{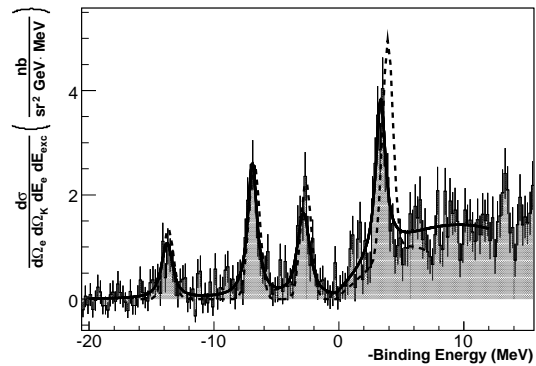}
\caption{The \lam{16}{N} binding-energy spectrum. The best fit using Voigt
functions (solid curve) and a theoretical prediction (dashed curve) are
superimposed on the data. See text for details.}
\label{oxygen2}
\end{figure}

 Figure~\ref{oxygen1} shows the binding-energy spectrum of \lam{16}{N}
for the full range of energy acceptance. The residual pion
contamination is $< 5$\%, uniformly distributed over the energy
spectrum (Fig.~\ref{oxygen1}). The shaded region shows the
 $(e,e') \otimes (e,K^+)$ random-coincidence background. 
The large broad peak observed at around 30 MeV is the mis-reconstructed 
binding energy due to the contribution from the hydrogen when the oxygen 
mass is used for the target mass in constructing the missing
mass. The excitation region of the \lam{16}{N} production is shown on 
the right side of the figure. The background is rather flat. However,
due to the acceptance of the detector, it decreases at the edges.
It is separately evaluated by plotting the data obtained for random
coincidences in a large energy window~\cite{iodice07} and fit
with a quadratic curve. No significant residual background is present
after subtraction.

Figure~\ref{oxygen2} shows the six-fold differential cross section
expressed in nb/(sr$^2$.GeV.MeV). The fit to the data has been made 
using Voigt functions, as described elsewhere~\cite{iodice07}. 
Four peaks are observed. The ground-state peak gives a $\Lambda$ 
separation energy of $B_\Lambda\!=\!13.76\pm 0.16$ MeV for the $1^-$ 
member of the ground-state doublet in \lam{16}{N}. Three more peaks are 
observed at binding energies of $6.93$, $2.84$, and $-3.34$ MeV. 
The measured energies, widths, and cross sections, after a radiative 
unfolding procedure, are given in Table~\ref{tab:results}.
Only statistical errors are reported for the measured cross-sections. 
Systematic errors, dominated by uncertainty in the target thickness, are
at the $20\%$ level.

The theoretical cross sections were obtained in the framework of
the distorted-wave impulse approximation (DWIA) \cite{sotona94} using
the Saclay-Lyon (SLA) model \cite{SLA} for the elementary
$p(e,e'K^+)\Lambda$ reaction. The ground state of $^{16}$O
is assumed to be a simple closed shell and the shell-model
wave functions for \lam{16}{N} are computed in a  particle-hole
model space. For the $p_N^{-1}s_\Lambda$ states, a multi-range
Gaussian (YNG) interaction ($r^2$ times a Gaussian for the tensor
interaction) is adjusted to reproduce the spectra of \lam{16}{O}
and \lam{15}{N} \cite{ukai08,millener08} when matrix elements are
evaluated with Woods-Saxon wave functions (with the $s_\Lambda$ orbit
bound by 13 MeV). This corresponds to fixing the conventional
parameters $\Delta$, $S_\Lambda$, $S_N$, and T for the spin-spin,
$\Lambda$-spin-orbit, nucleon-spin-orbit, and tensor components
of the $\Lambda N$ interaction at (in MeV)
\begin{equation}
\Delta= 0.330\quad S_\Lambda =-0.015\quad S_N = -0.350 \quad T =0.024 \; .
\label{eq:param}
\end{equation}
The effect of $\Lambda$-$\Sigma$ coupling is also taken into account
\cite{millener08}. The same YNG interaction is used for the 
$p_N^{-1}p_\Lambda$ states (with the $p_\Lambda$ orbits bound by 2.5 MeV). 
There are now 13 independent matrix elements compared with 5 for 
$p_N^{-1}s_\Lambda$. Most of the new matrix elements involve relative
$d$-wave states and are small. However, the relative $s$-wave and $p$-wave
central matrix elements no longer appear in a fixed combination. Here, the 
choice of the odd-state central interactions is influenced by data on
\lam{13}{C} \cite{kohri02,auerbach81}.

 The resulting energy spectra, dominant components of the wave functions, 
and calculated cross sections are shown in Table~\ref{tab:results}.
With respect to the calculations in \cite{iodice07}, an improved optical 
potential (with stronger kaon absorption) for the high $K^+$ energy in 
these experiments has been used. The four pronounced peaks in the 
energy spectrum are reproduced 
by the shell-model calculation. It is significant that the energy
separation between the two lowest peaks agrees very well with that
deduced from the theoretical centroids, and hence with the precise
$\gamma$-ray data \cite{ukai04,ukai08}. The excitation energies of the 
positive-parity states depend on the spacing of the $p_\Lambda$ and 
$s_\Lambda$ single-particle energies which have to be extrapolated 
from \lam{13}{C}~\cite{kohri02} and could therefore be uncertain by 
several hundred keV. The largest discrepancy between theory and 
experiment is in the position of the fourth peak. A firm prediction of 
the simple particle-hole model is that the gap between the third and 
fourth peaks should be slightly larger (6.5 MeV) than the underlying 
separation (6.324~MeV) of the p-hole states in $^{15}$N, in contrast 
to the observed splitting of $6.18$ MeV. It remains to be seen whether
a full $1\hbar\omega$ calculation, which incudes  $s_\Lambda$ states
coupled to positive-parity 1p2h states of the $^{15}$N core, leads to 
significant energy shifts and some fractionation of the $p_N^{-1}p_\Lambda$ 
strength.

 Historically, the first clear indication that the spin-orbit splitting 
of the $p_\Lambda$ orbits is very small came from the observation that 
the separation between the $0^+$ states is very close to the underlying 
core-state separation \cite{bruckner78}.
Table~\ref{tab:results} shows that the predicted major contributors to 
the third and fourth peaks involve both the $p_{1/2\Lambda}$ and 
$p_{3/2\Lambda}$ orbits and are indeed closely spaced for a 
$\Lambda$-spin-orbit strength corresponding to the very small value of 
$S_\Lambda$ in Eq.~(\ref{eq:param}). Increasing the spin-orbit strength 
would eventually cause an observable broadening of the fourth peak.
However, modest changes to any of the interaction parameters
would not lead to deviations from the observed four-peak structure.

 The $\Lambda$ separation energy $B_\Lambda\!=\!13.76\pm 0.16$ MeV 
obtained for the first peak is to be compared with $13.3\pm 0.4$ 
\cite{davis}, $13.4\pm 0.4$ \cite{tamura94,tamura}, and $12.4\pm 0.4$ 
\cite{hashtam06} for \lam{16}{O} from in-flight and stopped $K^-$
experiments.
The new $B_\Lambda$ value is an important quantity because {\em i) }
it depends on the average central $\Lambda N$ and perhaps $\Lambda NN$  
interactions in addition to the spin-dependence of the $\Lambda N$ 
interaction that primarily affects energy level spacings, {\em ii) } 
there are few emulsion events for the heavier p-shell hypernuclei 
and these events tend to have ambiguous interpretations~\cite{davis91}, 
and {\em iii) } the reactions involving the production of a $\Lambda$ 
from a neutron are more difficult to normalize. For example, the 
$(\pi^+,K^+)$ data  are normalized to 10.80 MeV for the $B_\Lambda$ of 
\lam{12}{C} \cite{hashtam06}, which differs considerably from the much 
better determined value of $11.37\pm 0.06$ MeV \cite{davis91} for the 
mirror hypernucleus \lam{12}{B}, and leads to $B_\Lambda$ values 
\cite{hashtam06} for \lamb{7}{Li}, \lamb{9}{Be}, \lam{10}{B}, and 
\lamb{13}{C} that are consistently lower than the emulsion values 
\cite{davis91}. Thus, the value of $12.4\pm 0.4$ MeV for \lam{16}{O}
may be an underestimate.

In summary, a high-quality \lam{16}{N} hypernuclear spectrum
has been obtained for the first time with sub-MeV energy resolution,
thus putting tighter restrictions on the spacings of the levels 
contributing to each peak.
The measured cross sections are in good agreement with the values
predicted using the SLA model and simple shell-model wave functions.
Most importantly, a $B_\Lambda$ value for \lam{16}{N} calibrated 
against the elementary $(e,e'K^+)$ reaction on hydrogen, has been 
obtained.

We acknowledge the Jefferson Lab Physics and Accelerator
Division staff for the outstanding efforts that made
this work possible.
This work was supported by U.S. DOE contract
DE-AC05-84ER40150, Mod. nr. 175,
under which the Southeastern Universities Research Association (SURA)
operates the Thomas Jefferson National Accelerator Facility,
 by the Italian Istituto Nazionale di Fisica
Nucleare and by the Grant Agency of the Czech Republic under grant No.
202/08/0984, by the french CEA and CNRS/IN2P3, and by the U.S. DOE under contracts, DE-AC02-06CH11357,
DE-FG02-99ER41110, and DE-AC02-98-CH10886, and by the U.S. National Science 
Foundation.

\end{document}